# An Inner SOCP Approximate Algorithm for Robust Adaptive Beamforming for General-Rank Signal Model

Yongwei Huang, *Senior Member, IEEE*, and Sergiy A. Vorobyov *Fellow, IEEE*

*Abstract*—The worst-case robust adaptive beamforming problem for general-rank signal model is considered. Its formulation is to maximize the worst-case signal-to-interference-plus-noise ratio (SINR), incorporating a positive semidefinite constraint on the actual covariance matrix of the desired signal. In the literature, semidefinite program (SDP) techniques, together with others, have been applied to approximately solve this problem. Herein an inner second-order cone program (SOCP) approximate algorithm is proposed to solve it. In particular, a sequence of SOCPs are constructed and solved, while the SOCPs have the nondecreasing optimal values and converge to a locally optimal value (it is in fact a globally optimal value through our extensive simulations). As a result, our algorithm does not use computationally heavy SDP relaxation technique. To validate our inner approximation results, simulation examples are presented, and they demonstrate the improved performance of the new robust beamformer in terms of the averaged cpu-time (indicating how fast the algorithms converge) in a high signal-to-noise region.

## I. Introduction

Robust adaptive beamforming is a powerful approach to significantly improve the array output signal-to-interference-plus-noise ratio (SINR) and other performance metrics such as mainlobe width, sidelobe levels, etc. [1]–[15]. Here the robustness typically means the ability of a method to perform well under any imperfect knowledge about the source, propagation media, and sensor array [8], [15]–[18]. There exists a number of robust adaptive beamforming approaches proposed for the scenario of a rank-one signal model (see [18] and reference therein). However, it is also of practical interest to consider a general-rank signal model, as the signal source often can be incoherently scattered, so that the design of robust adaptive beamforming for general-rank signal models becomes a must.

In [19], the authors have proposed an efficient robust adaptive beamforming method for general-rank source models, and a closed-form beamformer has been derived and good robustness capability has been demonstrated, ignoring however the positive semidefinite (PSD) constraint on the worst-case actual covariance for the desired signal. To fix the drawback, the authors of [20] have presented a new method to the robust adaptive beamforming, incorporating the PSD constraint. The resultant robust beamforming problem has been formulated by introducing a matrix decomposition (e.g. spectral or Cholesky type) of the presumed signal covariance and putting the error term into both of the matrices obtained from the decomposition. It turns out that the corresponding robust beamforming problem is a nonconvex quadratic program. However, the authors have proposed an algorithm for solving a semidefinite programming (SDP) problem in each iterative step, that finally enables them to obtain an approximate solution of the nonconvex quadratic program. In [21], two beamformers have been built in closed-form for the robust adaptive beamforming problem studied in [20], with the objective to lower the complexity of robust beamformers therein. The authors of [22] have proposed a method for solving the robust beamforming problem formulated in [20] using the SDP relaxation technique and bisection search. In each step of the bisection search, an SDP feasibility problem must be solved. In [23] and [24], the aforementioned beamforming problem has been treated as a difference-of-convex (DC) functions optimization problem, and a polynomial time DC (POTDC) approximate algorithm has been proposed, where an SDP problem had to be solved in every iterative step. Further, the authors have shown that under the condition that the error norm bound for the actual covariance of the desired signal is sufficiently small, the locally optimal solution output by the algorithm is indeed globally optimal.

In this paper, we study the robust adaptive beamforming problem, but avoid to use high complexity SDP problem solving-base iterations. Instead, we apply a second-order cone program (SOCP) to approximate the robust beamforming problem. By doing so, only a SOCP is solved at each iterative step, which has less computational cost. In particular, a sequence of SOCPs is constructed and solved, while the optimal values of the corresponding SOCPs are guaranteed to be nondecreasing and to converge to a locally optimal value. Moreover, in our extensive simulations, we find out that our approximate algorithm converges to a globally optimal solution. Since every SOCP is a restriction of the robust beamforming problem, the approximate algorithm is said to be an inner approximation. To the best of our knowledge, this is the first time to propose a SOCP approximate algorithm for the general-rank robust beamforming problem under consideration (without utilizing the SDP relaxation technique).

## II. Signal Model and Problem Formulation

The array output at time instance $t$ is expressed as

$$x(t) = \boldsymbol{w}^H \boldsymbol{y}(t)$$

where $\boldsymbol{w}$ is the $N \times 1$ complex weight (beamforming) vector, $\boldsymbol{y}(t)$ is the $N \times 1$ complex snapshot vector of array observa-

Y. Huang is with School of Information Engineering, Guangdong University of Technology, University Town, Guangzhou, Guangdong, 510006, China. Email: ywhuang@gdut.edu.cn.

S. A. Vorobyov is with Department of Signal Processing and Acoustics, School of Electrical Engineering, Aalto University, Konemiehentie 2, 02150 Espoo, Finland. Email: svor@ieee.org.

tions, $N$ is the number of sensors in the array, and $(\cdot)^H$ stands for Hermitian transpose. The observation vector is given by

$$\boldsymbol{y}(t) = \boldsymbol{s}(t) + \boldsymbol{i}(t) + \boldsymbol{n}(t) \quad (1)$$

where $\boldsymbol{s}(t)$, $\boldsymbol{i}(t)$, and $\boldsymbol{n}(t)$ are the statistically independent components of the desired signal, interference, and noise, respectively. The output SINR of the beamformer is written as

$$\text{SINR} = \frac{\boldsymbol{w}^H \boldsymbol{R}_s \boldsymbol{w}}{\boldsymbol{w}^H \boldsymbol{R}_{i+n} \boldsymbol{w}} \quad (2)$$

where $\boldsymbol{R}_s \triangleq \mathsf{E}[\boldsymbol{s}(t)\boldsymbol{s}^H(t)]$ is the desired signal covariance matrix and $\boldsymbol{R}_{i+n} \triangleq \mathsf{E}[(\boldsymbol{i}(t)+\boldsymbol{n}(t))(\boldsymbol{i}(t)+\boldsymbol{n}(t))^H]$ is the interference-plus-noise covariance matrix. Matrix $\boldsymbol{R}_s$ herein can be of rank one or higher, i.e., $\text{Rank}(\boldsymbol{R}_s) \in \{1, \ldots, N\}$. The rank-one $\boldsymbol{R}_s$ corresponds to the case of the point source (see [16], [18]–[20]), and we herein are interested in the general-rank case.

In applications, the matrix $\boldsymbol{R}_{i+n}$ is typically unavailable. Thus, the sample covariance of the date covariance matrix $\boldsymbol{R} \triangleq \mathsf{E}[\boldsymbol{y}(t)\boldsymbol{y}^H(t)]$ is used, and it is expressed as

$$\hat{\boldsymbol{R}} = \frac{1}{T} \sum_{t=1}^{T} \boldsymbol{y}(t)\boldsymbol{y}^H(t) \quad (3)$$

where $T$ stands for the number of training snapshots. On the other hand, the actual covariance matrix $\boldsymbol{R}_s$ is only known incompletely and imperfectly. The beamvector obtained by maximizing the SINR (2) with $\boldsymbol{R}_s$ and $\boldsymbol{R}_{i+n}$ replaced respectively by $\hat{\boldsymbol{R}}_s$ (a presumed covariance of $\boldsymbol{R}_s$) and $\hat{\boldsymbol{R}}$ can however lead to poor performance of the sensor array. Therefore, in order to improve the beamformer performance, robust adaptive beamforming for general-rank signal model has been considered. Several works have addressed the problem (see [19]–[24] and references therein) in the last two decades. Among robust adaptive beamforming problem formulations, the following problem maximizing the worst-case SINR is the most popular

$$\underset{\boldsymbol{w} \neq \boldsymbol{0}}{\text{maximize}} \ \underset{\boldsymbol{\Delta}_1 \in \mathcal{B}_1, \boldsymbol{\Delta}_2 \in \mathcal{B}_2}{\min} \frac{\boldsymbol{w}^H(\hat{\boldsymbol{R}}_s + \boldsymbol{\Delta}_2)\boldsymbol{w}}{\boldsymbol{w}^H(\hat{\boldsymbol{R}} + \boldsymbol{\Delta}_1)\boldsymbol{w}} \quad (4)$$

where the uncertainty sets $\mathcal{B}_1$ and $\mathcal{B}_2$ are given by

$$\mathcal{B}_1 = \{\boldsymbol{\Delta}_1 \in \mathbb{C}^{N \times N} \mid \|\boldsymbol{\Delta}_1\| \leq \gamma, \hat{\boldsymbol{R}} + \boldsymbol{\Delta}_1 \succeq \boldsymbol{0}\} \quad (5)$$

and

$$\mathcal{B}_2 = \{\boldsymbol{\Delta}_2 \in \mathbb{C}^{N \times N} \mid \|\boldsymbol{\Delta}_2\| \leq \epsilon, \hat{\boldsymbol{R}}_s + \boldsymbol{\Delta}_2 \succeq \boldsymbol{0}\} \quad (6)$$

respectively. Throughout the paper, the matrix norm $\|\cdot\|$ is assumed to be the Frobenius norm, while a vector norm is Euclidian norm.

Since $\boldsymbol{\Delta}_1$ and $\boldsymbol{\Delta}_2$ are separable, (4) can be recast into

$$\underset{\boldsymbol{w} \neq \boldsymbol{0}}{\text{maximize}} \ \frac{\underset{\boldsymbol{\Delta}_2 \in \mathcal{B}_2}{\min} \boldsymbol{w}^H(\hat{\boldsymbol{R}}_s + \boldsymbol{\Delta}_2)\boldsymbol{w}}{\underset{\boldsymbol{\Delta}_1 \in \mathcal{B}_1}{\max} \boldsymbol{w}^H(\hat{\boldsymbol{R}} + \boldsymbol{\Delta}_1)\boldsymbol{w}}. \quad (7)$$

Observe that for $\boldsymbol{\Delta}_1 \in \mathcal{B}_1$, it follows that

$$\boldsymbol{w}^H \boldsymbol{\Delta}_1 \boldsymbol{w} = \text{tr}(\boldsymbol{\Delta}_1 \boldsymbol{w}\boldsymbol{w}^H) \leq \|\boldsymbol{\Delta}_1\| \cdot \|\boldsymbol{w}\boldsymbol{w}^H\| \leq \gamma \|\boldsymbol{w}\|^2$$

and the equality holds when $\boldsymbol{\Delta}_1 = \gamma \boldsymbol{w}\boldsymbol{w}^H / \|\boldsymbol{w}\|^2 \in \mathcal{B}_1$. Therefore, the solution to the problem in the denominator of the objective function of (7) can be found in closed-form as

$$\boldsymbol{w}^H(\hat{\boldsymbol{R}} + \gamma \boldsymbol{I})\boldsymbol{w}. \quad (8)$$

Thus, (7) can be reexpressed as

$$\underset{\boldsymbol{w} \neq \boldsymbol{0}}{\text{maximize}} \ \frac{\underset{\boldsymbol{\Delta}_2 \in \mathcal{B}_2}{\min} \boldsymbol{w}^H(\hat{\boldsymbol{R}}_s + \boldsymbol{\Delta}_2)\boldsymbol{w}}{\boldsymbol{w}^H(\hat{\boldsymbol{R}} + \gamma \boldsymbol{I})\boldsymbol{w}}, \quad (9)$$

which is equivalent to the following optimization problem

$$\begin{aligned} \underset{\boldsymbol{w}}{\text{maximize}} \quad & \underset{\boldsymbol{\Delta}_2 \in \mathcal{B}_2}{\min} \boldsymbol{w}^H(\hat{\boldsymbol{R}}_s + \boldsymbol{\Delta}_2)\boldsymbol{w} \\ \text{subject to} \quad & \boldsymbol{w}^H(\hat{\boldsymbol{R}} + \gamma \boldsymbol{I})\boldsymbol{w} \leq 1, \end{aligned} \quad (10)$$

in the sense that (9) and (10) share the same optimal value and if $\boldsymbol{w}^\star$ solves (10), then it is optimal for (9) too.

In order to incorporate the PSD constraint $\hat{\boldsymbol{R}}_s + \boldsymbol{\Delta}_2 \succeq \boldsymbol{0}$ over $\mathcal{B}_2$ into the objective function in (10), the following robust adaptive beamforming problem is considered (cf. [20]–[24])

$$\begin{aligned} \underset{\boldsymbol{w}}{\text{maximize}} \quad & \underset{\boldsymbol{\Delta} \in \mathcal{B}}{\min} \boldsymbol{w}^H(\boldsymbol{Q}+\boldsymbol{\Delta})^H(\boldsymbol{Q}+\boldsymbol{\Delta})\boldsymbol{w} \\ \text{subject to} \quad & \boldsymbol{w}^H(\hat{\boldsymbol{R}} + \gamma \boldsymbol{I})\boldsymbol{w} \leq 1 \end{aligned} \quad (11)$$

where $\boldsymbol{Q}^H \boldsymbol{Q} = \hat{\boldsymbol{R}}_s$, $\boldsymbol{Q} \in \mathbb{C}^{M \times N}$, $N \geq M = \text{Rank}(\boldsymbol{R}_s)$, and the norm of the distortion $\boldsymbol{\Delta}$ is bounded by a given constant $\eta$:

$$\mathcal{B} = \{\boldsymbol{\Delta} \in \mathbb{C}^{M \times N} \mid \|\boldsymbol{\Delta}\| \leq \eta\}. \quad (12)$$

Herein we focus on how to solve the beamforming problem (11), and present a new algorithm which we coin as the inner SOCP approximate algorithm.

It is known that [24], [25]

$$\underset{\boldsymbol{\Delta} \in \mathcal{B}}{\min} \boldsymbol{w}^H(\boldsymbol{Q}+\boldsymbol{\Delta})^H(\boldsymbol{Q}+\boldsymbol{\Delta})\boldsymbol{w} = (\max\{\|\boldsymbol{Q}\boldsymbol{w}\| - \eta\|\boldsymbol{w}\|, 0\})^2.$$

Considering that the objective function of (11) is positive, we rewrite (11) as the following equivalent problem

$$\begin{aligned} \underset{\boldsymbol{w}}{\text{maximize}} \quad & \|\boldsymbol{Q}\boldsymbol{w}\| - \eta\|\boldsymbol{w}\| \\ \text{subject to} \quad & \boldsymbol{w}^H \hat{\boldsymbol{R}} \boldsymbol{w} + \gamma\|\boldsymbol{w}\|^2 \leq 1. \end{aligned} \quad (13)$$

For comparison, in [20], the authors have studied the following problem:

$$\begin{aligned} \underset{\boldsymbol{w}}{\text{minimize}} \quad & \boldsymbol{w}^H \hat{\boldsymbol{R}} \boldsymbol{w} + \gamma\|\boldsymbol{w}\|^2 \\ \text{subject to} \quad & \|\boldsymbol{Q}\boldsymbol{w}\| - \eta\|\boldsymbol{w}\| \geq 1. \end{aligned} \quad (14)$$

Notice that both the inequality constraints in (13) and (14) can be replaced with equality constraints, without loss of anything.

We show next that the problems (13) and (14) are equivalent to each other.

**Proposition II.1** *Problems* (13) *and* (14) *are equivalent to each other in the sense that if $\boldsymbol{w}^\star$ and $v^\star$ are respectively an optimal solution and the optimal value of* (14), *then $\boldsymbol{w}^\star/\sqrt{v^\star}$ and $1/\sqrt{v^\star}$ are correspondingly an optimal solution and the optimal value of* (13). *Conversely, if $\boldsymbol{w}^\star$ and $v^\star$ are respectively an optimal solution and the optimal value of* (13), *then $\boldsymbol{w}^\star/v^\star$ and $1/v^{\star 2}$ are correspondingly an optimal solution and the optimal value of* (14).

*Proof:* Suppose that $w^\star$ and $v^\star$ are respectively an optimal solution and the optimal value of (14). It follows that

$$(w^\star/\sqrt{v^\star})^H(\hat{R}+\gamma I)(w^\star/\sqrt{v^\star})=1$$

and

$$\|Qw^\star\|-\eta\|w^\star\|=1.$$

Suppose that $w^\star/\sqrt{v^\star}$ is not the optimal solution of (13), and there is an optimal solution $\tilde{w}$ such that

$$\|Q\tilde{w}\|-\eta\|\tilde{w}\| > \frac{\|Qw^\star\|-\eta\|w^\star\|}{\sqrt{v^\star}} = \frac{1}{\sqrt{v^\star}},$$

and

$$\tilde{w}^H(\hat{R}+\gamma I)\tilde{w}=1,$$

which implies respectively that

$$\|Q(\sqrt{v^\star}\tilde{w})\|-\eta\|\sqrt{v^\star}\tilde{w}\| > 1, \quad (15)$$

and

$$(\sqrt{v^\star}\tilde{w})^H(\hat{R}+\gamma I)(\sqrt{v^\star}\tilde{w})=v^\star.$$

This means that $\sqrt{v^\star}\tilde{w}$ is optimal for (14), but with respect to (15), it is impossible since (15) shall hold with equality. Then a contradiction occurs, and thereby we conclude that $w^\star/\sqrt{v^\star}$ is optimal for (13) with the optimal value $1/\sqrt{v^\star}$.

The proof for the converse implication is similar to the above proof, and thus it is omitted. ∎

Given the Proposition, we can solve (14) in order to get an optimal solution of (11).

## III. AN INNER SOCP APPROXIMATE PROCEDURE TO SOLVE (14)

We design here an inner SOCP approximate algorithm for nonconvex problem (14).

To begin, problem (14) is reformulated equivalently into

$$\begin{aligned}\underset{w,t}{\text{minimize}} \quad & w^H\hat{R}w+\gamma\|w\|^2 \\ \text{subject to} \quad & t-\eta\|w\| \geq 1 \\ & \|Qw\| \geq t.\end{aligned} \quad (16)$$

Indeed, it can be verified that the optimal solutions of one problem are always feasible for the other problem. Note that the first constraint in (16) is a second-order cone (SOC) constraint, but the second one is a nonconvex constraint.

Nevertheless, we can build a convex approximation for the nonconvex constraint, and design based on it an algorithm to solve (16). Towards this end, observe that

$$\|Qw\| \geq \frac{|w_0^H Q^H Qw|}{\|Qw_0\|} \geq \frac{\Re(w_0^H Q^H Qw)}{\|Qw_0\|},$$

where $w_0$ is an initial point. Therefore,

$$\frac{\Re(w_0^H Q^H Qw)}{\|Qw_0\|} \geq t$$

implies that

$$\|Qw\| \geq t$$

and we establish the following SOCP problem instead of SDP

$$\begin{aligned}\underset{w,t}{\text{minimize}} \quad & w^H\hat{R}w+\gamma\|w\|^2 \\ \text{subject to} \quad & t-\eta\|w\| \geq 1 \\ & \frac{\Re(w_0^H Q^H Qw)}{\|Qw_0\|} \geq t\end{aligned} \quad (17)$$

which has a smaller feasible set and its optimal value is bigger than that of (16). Here $\Re(\cdot)$ stands for a real part of a complex number. In other words, problem (17) is a restriction of (16).

We wish to design a sequence of problems in the form of (17) such that the optimal value of each consecutive problem is closer to a locally optimal value (but in our extensive simulations, it converges to the globally optimal value).

Suppose $(w_1, t_1)$ is an optimal solution for (17) and the optimal value is $v_1$. Then we construct the following SOCP problem

$$\begin{aligned}\underset{w,t}{\text{minimize}} \quad & w^H\hat{R}w+\gamma\|w\|^2 \\ \text{subject to} \quad & t-\eta\|w\| \geq 1, \\ & \frac{\Re(w_1^H Q^H Qw)}{\|Qw_1\|} \geq t.\end{aligned} \quad (18)$$

Clearly it is also a restriction of (16) since the last constraint implies that $\|Qw\| \geq t$. Solving (18), we obtain an optimal solution $(w_2, t_2)$ and the optimal value $v_2$. Then the following proposition about a relationship between optimal values $v_1$ and $v_2$ can be proven.

**Proposition III.1** *It holds that $v_2 \leq v_1$.*

*Proof:* It follows from the second constraint of (17) that

$$\|Qw_1\| \geq \frac{\Re(w_0^H Q^H Qw_1)}{\|Qw_0\|} \geq t_1.$$

It can be easily checked that the optimal solution $(w_1, t_1)$ is feasible for (18). Therefore, we have $v_2 \leq v_1$. ∎

Similarly, we can construct a sequence of SOCP problems, such that the optimal values comply with

$$v_1 \geq v_2 \geq v_3 \geq \ldots.$$

Therefore, we summarize the inner SOCP approximate algorithm as shown in Algorithm 1.

---

**Algorithm 1** Inner SOCP approximate algorithm for (14)
---
**Input:** $\hat{R}$, $Q$, $\gamma$, $\eta$, $\xi$;
**Output:** A solution $w^\star$ for problem (14);
1: Suppose that $w_0$ is an initial feasible point; set $k=0$ and $v_0 = w_0^H \hat{R} w_0 + \gamma\|w_0\|^2$;
2: **do**
3:     solve the following SOCP:

$$\begin{aligned}\underset{w,t}{\text{minimize}} \quad & w^H\hat{R}w+\gamma\|w\|^2 \\ \text{subject to} \quad & t-1 \geq \eta\|w\|, \\ & \frac{\Re(w_k^H Q^H Qw)}{\|Qw_k\|} \geq t,\end{aligned} \quad (19)$$

    obtaining solution $w^\star$ and optimal value $v^\star$;
4:     $k:=k+1$;
5:     $w_k = w^\star$; $v_k = v^\star$;
6: **until** $v_{k-1} - v_k \leq \xi$

---

It is worth observing that every SOCP in Algorithm 1 is a restriction of (16). We remark that the computational burden includes solving SOCP (19) in each iteration, and the worst-case complexity of solving the SOCP is in the order of $O(N^3)$ (see e.g. [26, page 309]). We also remark that in the approximate algorithm proposed in [24], an SDP is solved in

every iteration, which has higher computational cost. In fact, therein (14) is recast into

$$\begin{aligned}
\underset{\boldsymbol{W},\alpha}{\text{minimize}} \quad & \text{tr}((\hat{\boldsymbol{R}}+\gamma\boldsymbol{I})\boldsymbol{W}) \\
\text{subject to} \quad & \text{tr}(\boldsymbol{Q}^H\boldsymbol{Q}\boldsymbol{W}) = \alpha \\
& \eta^2\text{tr}\,\boldsymbol{W} \leq (\sqrt{\alpha}-1)^2 \\
& \boldsymbol{W} \succeq \boldsymbol{0},\ \text{rank}(\boldsymbol{W})=1,\ \theta_1 \leq \alpha \leq \theta_2.
\end{aligned} \quad (20)$$

Employing the approximation $\sqrt{\alpha} \approx \sqrt{\alpha_0} + (\alpha-\alpha_0)/(2\sqrt{\alpha_0})$ and the conventional SDP relaxation technique, (20) is relaxed into the following SDP

$$\begin{aligned}
\underset{\boldsymbol{W},\alpha}{\text{minimize}} \quad & \text{tr}((\hat{\boldsymbol{R}}+\gamma\boldsymbol{I})\boldsymbol{W}) \\
\text{subject to} \quad & \text{tr}(\boldsymbol{Q}^H\boldsymbol{Q}\boldsymbol{W}) = \alpha \\
& \eta^2\text{tr}\,\boldsymbol{W} \leq \alpha(1-1/\sqrt{\alpha_0}) + 1 - \sqrt{\alpha_0} \\
& \boldsymbol{W} \succeq \boldsymbol{0},\ \theta_1 \leq \alpha \leq \theta_2.
\end{aligned} \quad (21)$$

Then the algorithm searches for the best $\alpha$, and in each iteration, solves the corresponding SDP (21) for finding optimal $\boldsymbol{W}$. At the end, $\boldsymbol{W}$ [1] for the best $\alpha$ gives a suboptimal solution of (20), which is proven to be globally optimal under some conditions. Interestingly, the optimal value of the SDP relaxation of (20) (dropping the rank-one constraint) can serve as a lower bound of the robust beamforming problem (14) (a benchmark, as will be seen in simulation examples).

Finally, we point out that (13) can be solved directly, that is, it can be approximated by the following SOCP problem

$$\begin{aligned}
\underset{\boldsymbol{w},s,t}{\text{maximize}} \quad & s \\
\text{subject to} \quad & t - \eta\|\boldsymbol{w}\| \geq s, \\
& \frac{\Re(\boldsymbol{w}_k^H\boldsymbol{Q}^H\boldsymbol{Q}\boldsymbol{w})}{\|\boldsymbol{Q}\boldsymbol{w}_k\|} \geq t, \\
& \boldsymbol{w}^H\hat{\boldsymbol{R}}\boldsymbol{w} + \gamma\|\boldsymbol{w}\|^2 \leq 1
\end{aligned} \quad (22)$$

provided that an initial point $\boldsymbol{w}_0$ is a prefix. However we choose to solve (14) for the reason that it is easier to compare this our work with the existing works.

## IV. SIMULATION RESULTS

We consider a uniform linear array of 10 omni-directional antenna elements with the inter-element spacing of half wavelength (i.e. $N=10$). The power of additive noise in every antenna is assumed to be 0 dB. There is an interferer with the interference-to-noise ratio (INR) of 20 dB. Suppose that the desired signal and the interferer are locally incoherently scattered with Gaussian and uniform angular power densities with central angles of $30°$ and $10°$, respectively. The angular spreads of the desired signal and the interferer are assumed to be $4°$ and $10°$, respectively. The angular power density of the presumed signal is Gaussian with central angle $34°$ and angular spread $6°$. The diagonal loading parameter $\gamma = 0.1\|\hat{\boldsymbol{R}}\|$ and the error norm bound $\eta = 0.5\sqrt{\text{tr}(\hat{\boldsymbol{R}}_s)}$ are used, where $\text{tr}(\cdot)$ denotes the matrix trace. The iteration termination threshold $\xi$ is equal to $10^{-8}$. To find a lower-bound on the optimal value of SDP relaxation of (20), the interval $[\theta_1, \theta_2]$ is divided into 100 subsectors. All results are averaged over 100 simulation runs.

---
[1] It is easily verified that $\boldsymbol{W}$ can be always of rank one, see [27]. More generally, for any given $\alpha$, the SDP (21) admits a rank-one solution.

Fig. 1 shows the required CPU-time by our Algorithm 1 herein ("New beamformer") and the algorithm proposed in [24]) ("K-V beamformer"). It can be seen that the averaged CPU-time of our beamformer is less than that by [24], especially at the high SNR region (our computer has a processor of Inter Core i7-6650U and a 16 GB RAM). Fig. 2 depicts the approximate optimal values by our algorithm and the one in [24], as well as the optimal value of (14), or equivalently (20) (termed as "Lower-Bound"). It can be seen that the three curves coincide, meaning that both our approximate algorithm and the approach in [24] achieve the optimal value of the original problem (14).

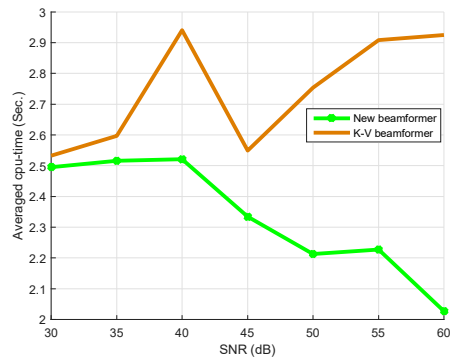

Fig. 1. Averaged cpu-time versus SNR, with INR=20 dB and $T=50$

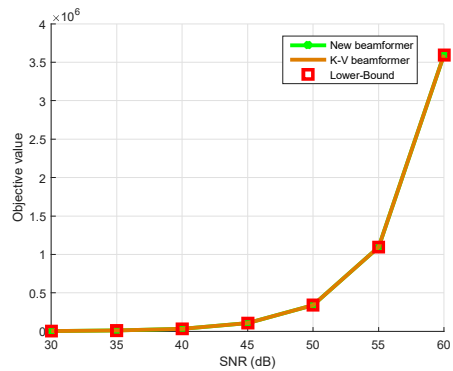

Fig. 2. Objective function value of the problem (14) (or equivalently (20)) versus SNR, with INR=20 dB and $T=50$

## V. CONCLUSION

We have considered the robust adaptive beamforming problem for general-rank signal models incorporating one PSD constraint over the actual covariance matrix of the desired signal. Unlike solving the problem by the SDP relaxation in every iterative step, we have developed the inner SOCP approximate algorithm. In particular, a sequence of SOCPs have been constructed and solved, where the optimal values of the SOCPs are nondecreasing and converge to an optimal value of the problem considerd. Since only solving an SOCP is required in iterative steps, the total computational complexity is lower than that of existing algorithms involving solving SDPs. The improved performance of the proposed robust beamformer has been demonstrated by simulations in terms of the averaged CPU-time required for the algorithms, which indicate how fast the algorithms are.